\documentclass[10pt,aps,prd,twocolumn,showpacs,showkeys]{revtex4}

\usepackage{amsmath,amssymb}
  \usepackage{epsfig}
\usepackage{graphicx}% Include figure files
\usepackage{bm}
\usepackage{amsmath}
\newcommand{\beq}{\begin{equation}}
\newcommand{\eeq}{\end{equation}}
\newcommand{\beqa}{\begin{eqnarray}}
\newcommand{\eeqa}{\end{eqnarray}}
\newcommand{\beqar}{\begin{eqnarray*}}
\newcommand{\eeqar}{\end{eqnarray*}}

\newcommand\munu{\ensuremath{{\mu\nu}}}

\begin{document}

\title{Regular black holes: Electrically charged solutions,
Reissner-Nordstr\"om outside a de Sitter core}
%\title{Regular Reissner-Nordstr\"om black holes}

\author{Jos\'e P. S. Lemos}
\affiliation {Centro Multidisciplinar de Astrof\'{\i}sica -- CENTRA,
Departamento de F\'{\i}sica,  Instituto Superior T\'ecnico - IST,
Universidade T\'ecnica de Lisboa - UTL,
Av. Rovisco Pais 1, 1049-001 Lisboa, Portugal
\\Email: joselemos@ist.utl.pt}

\author {Vilson T. Zanchin}
\affiliation {Centro de Ci\^encias Naturais e Humanas, Universidade
Federal
do ABC,  Rua Santa Ad\'elia, 166, 09210-170, Santo Andr\'e, SP,
Brazil,
\\\&\\
Coordenadoria de Astronomia e Astrof\'{\i}sica, Observat\'orio
Nacional-MCT, Rua General Jos\'e Cristino 77, 20921-400 Rio de Janeiro,
Brazil\\ Email: zanchin@ufabc.edu.br}

%\date{today}

\begin{abstract}

To have the correct picture of a black hole as a whole it is of
crucial importance to understand its interior.  The singularities that
lurk inside the horizon of the usual Kerr-Newman family of black hole
solutions signal an endpoint to the physical laws and as such should
be substituted in one way or another.  A proposal that has been around
for sometime, is to replace the singular region of the spacetime by a
region containing some form of matter or
false vacuum configuration that can also cohabit with the black
hole interior. Black holes without singularities are called regular
black holes. In the present work regular black hole solutions are
found within general relativity coupled to Maxwell's electromagnetism
and charged matter. We show that there are objects which correspond to
regular charged black holes, whose interior region is de Sitter, whose
exterior region is Reissner-Nordstr\"om, and the boundary between both
regions is made of an electrically charged spherically symmetric
coat. There are several type of solutions: regular nonextremal black
holes with a null matter boundary, regular nonextremal black holes
with a timelike matter boundary, regular extremal black holes with a
timelike matter boundary, and regular overcharged stars with a
timelike matter boundary.  The main physical and geometrical
properties of such charged regular solutions are analysed.


\end{abstract}

%\begin{keyword}
%\end{keyword}

%\keywords{Quasiblack holes, Black holes,
%Classical theories of gravity}

\maketitle

%\end{frontmatter}

\date{today}
\section {Introduction}
\label{sec-introd}

\subsection {Solutions of Einstein's equation and black holes}
\label{subsec-solbhs}

Finding solutions of Einstein's equation is an important branch of
general relativity.  The equation takes the form, $ G_{\mu\nu}=8\pi\,
\tau_{\mu\nu} $, where $G_{\mu\nu}$ is the Einstein tensor,
constructed from second derivatives of the metric tensor $g_{\mu\nu}$,
$\tau_{\mu\nu}$ is the stress-energy tensor of the matter and fields,
$\mu,\nu$ are spacetime indices $\mu,\nu=0,1,2,3$, and $G=1\,,c=1$.
Arbitrarily chosen spacetimes with the corresponding metrics
$g_{\mu\nu}$ usually give rise to an Einstein tensor which in turn
corresponds to an unphysical stress-tensor, i.e., to matter which is
of no interest. Thus, finding solutions of Einstein's equation is in
general a non-trivial task.  The efforts to overtake the difficulties
are displayed in the exact solutions book \cite{stephanietal}, where
several methods of solving Einstein's equation are given.  To
facilitate the work, instead of having a continuous solution
throughout spacetime, one can find solutions for two regions, an
interior region, and an exterior region, and then matches these
regions through a smooth junction, a boundary surface \cite{israel}.
One can also opt for a more drastic junction between both regions
where a surface layer, i.e., a thin shell, is needed \cite{israel}.
Usually the junctions, or solderings, are through
timelike surfaces, as in a surface
of a star. One can also extend the formalism of \cite{israel} to
spacelike surfaces.  For a lightlike surface (also called null
surface) a special formalism can be developed, as in, e.g.,
\cite{barrabes_hogan}.

A well known vacuum ($\tau_{\mu\nu}=0$) solution of Einstein's
equation is the Schwarzschild black hole.  This spherically symmetric
solution has an event horizon at a coordinate radius $r_{\rm h}=2m$,
where $m$ is the mass of the black hole (see, e.g., \cite{mtw}).  In
its full form the solution represents a wormhole, with its two phases,
the white hole and the black hole, both phases harboring singularities
and connecting two asymptotically flat universes \cite{mtw}.  In its
amputated form, the solution can represent a black hole shielding a
singularity and with one asymptotically flat region, the black hole
being formed from the collapse of a star or some other lump of matter
(see, e.g., \cite{mtw}).  When, besides a mass $m$, one includes
electrical charge $q$, the Reissner-Nordstr\"om solution is obtained
(see, e.g., \cite{mtw}). The inclusion of angular momentum $J$ gives
the Kerr solution, and the inclusion of the three parameters ($m,J,q$)
is the Kerr-Newman solution, yielding the Kerr-Newman family of black
holes within general relativity (see \cite{mtw}, see also
\cite{grifithspodolsky2009}).

The outside of a black hole is visible. Nowadays potent telescopes and
detectors watch with ease what is going on in jets and phenomena
powered out by black holes. Moreover, the outside of a black hole is
well known classically (see, e.g., \cite{stewartwalker}). Thus, from
the outside, there is astrophysical evidence and theoretical
consistency, within the classical framework of general relativity, for
the existence of black holes. Quantically, however, black holes still
pose problems for the outside. These problems are related to the
Hawking radiation and the Bekenstein-Hawking entropy (see, e.g.,
\cite{frolnov}).  Although their solution is not in hand, the quantum
outside problems are well posed and delineated.

The inside of a black hole, on the other hand, is another story, it is
not known at all.  By definition the black hole inside is hidden, it
encloses a mysterious unknown.  The understanding of the inside of a
black hole is one of the outstanding problems in gravitational theory.
The Schwarzschild solution describes the black hole inside as an ever
moving spacetime that ends on an all encompassing spacelike
singularity. The Reissner-Nordstr\"om solution also has an ever moving
inward spacetime that, instead, ends on a Cauchy horizon which can
then be cruised into a region where a singularity can be seen but
avoided.  The Kerr and the Kerr-Newman solutions have analogous
properties to the Reissner-Nordstr\"om solution.  The event horizon
thus, for this class of black hole solutions, harbors a
singularity. What is a singularity?  The singularity theorems
\cite{penrose65,hawkingpenrose,hawkingbook1973,penrose78} do not tell
what a singularity is. Through the imposition of some precise physical
conditions the theorems only prove generically that singularities are
inevitable.  But those precise physical conditions might not be upheld
in the situations they are to be used, turning the theorems
useless in such        circumstances.
The existence of a singularity, by its very definition,
means spacetime ceases to exist signaling a failure of the physical
laws.
So, if physical laws do exist at those extreme conditions,
singularities should be substituted by some other object in a more
encompassing theory.  The extreme conditions, in one form or another,
that may exist at a singularity, imply that one should resort to
quantum gravity. Singularities are certainly objects
to be resolved in the realm of quantum gravity \cite{wheeler}.

Since there is no definite quantum gravity yet, a line of work to
understand the inside of a black hole and resolve
its singularity, is to study
classical or semiclassical black holes, with regular, i.e.,
nonsingular, properties. These type of black holes can be motivated by
quantum arguments.  In this way, there has been a trend to invent
regular black hole solutions with special matter cores that would
substitute the true singularities of the Schwarzschild,
Reissner-Nordstr\"om, Kerr, and Kerr-Newman black holes.

\subsection{Early considerations}
\label{subsec-eraly}

Indeed, Sakharov \cite{sakharov} and Gliner \cite{gliner1966} in 1966
proposed that singularities, such as cosmological singularities, could
be avoided by matter at superhigh densities with an inflationary
equation of state, i.e., with a de Sitter core, in which the equation
of state between the pressure $p$ and the energy density $\rho_{\rm
matter}$ is $p=-\rho_{\rm matter}$, or, equivalently, the matter
stress-energy tensor $T_{\mu\nu}$ takes a lambda term or false vacuum
form $T_{\mu\nu}=\Lambda g_{\mu\nu}$, $\Lambda$ being the cosmological
constant.  This led to the idea that a spacetime filled with false
vacuum inside a black hole horizon could provide a description of the
final state of gravitational collapse.  Furthermore, Zel'dovich
\cite{zel68} proposed afterwards that such a $T_{\mu\nu}$ could arise
naturally as a result of vacuum polarization processes in
gravitational interactions. These considerations may indicate that an
unlimited increase of spacetime curvature during a collapse process
can lead to the halt of the collapse itself if quantum fluctuations
dominate the process, putting ultimately an upper bound to the value
of the curvature and obliging the formation of a central core.

\subsection {The Bardeen regular black hole}
\label{subsec-bard}

Bardeen in 1968 \cite{bardeen1968} realized concretely the
idea of a central matter core, by
proposing a solution of Einstein's equation in which there is a
black hole with horizons but without a singularity, the first regular
black hole.  The matter field content was a kind of magnetic matter
field, yielding a modification of the Reissner-Nordstr\"om metric. But
near the center the solution tended to a de Sitter core solution.  All
the subsequent regular black hole solutions are based on Bardeen's
proposal, although there has been a tremendous development on the
implementation and on the analysis of the properties of regular black
hole solutions.

\subsection{Other regular black holes}
\label{subsec-other}

A useful way to classify the regular black hole solutions is through
the type of junctions needed. If there is no junction, the solution is
a continuous solution throughout spacetime.  If there are two simple
regions, the solutions have boundary surfaces joining the two regions.
In more drastic cases the solutions have surface layers, i.e., thin
shells joining the two regions.

\subsubsection{Solutions with continuous fields}

Based on a previous work on how to avoid cosmological singularities
\cite{gliner1975}, Dymnikova proposed in 1992 \cite{dy92} a black hole
model in which the core is de Sitter and gives way in a smooth manner
into a Schwarzschild solution, with Cauchy and event horizons
somewhere in between.  Several subsequent works developing this idea
followed \cite{dy96,dy00,dy01,dy03,dy04,dy05,dy10}, see also
\cite{gliner1998}.  Next, Ay\'on-Beato and Garc\'\i a invoked nonlinear
fields and sources to generate from first principles the Bardeen model
as a nonlinear magnetic monopole \cite{ab00}, and also found a four
parameter solution \cite{ab05}.  In addition,
they also attempted to derive regular
black holes from nonlinear electric fields \cite{ab98}, see also
\cite{brcritic,br01}, and see \cite{mat04} for the extremal limit of
the solutions. Bronnikov and collaborators also produced several
regular black holes in which the source are fields permeating the
whole spacetime, the core is an expanding universe with de
Sitter asymptotics and the exterior outer region tends to
Schwarzschild \cite{br06,br071,br072}.  In \cite{mat08} a development
along the same lines can be found.
Regular black holes in quadratic gravity
have also been discussed \cite{mat06}.

\subsubsection{Solutions with boundary surfaces}

Another useful way to construct regular black holes
is by filling the inner space with
matter up to a certain surface and then make a smooth
junction, through a boundary surface, to
the Schwarzschild solution as was done in
\cite{mars1996,magli,lake2005}, and in a more general
setting in \cite{elizalde2002}.
In the junction to the Schwarzschild case this
junction is made through a spacelike surface, rather than an usual
timelike surface. This means that, for some parametrization, the
junction exists at a single instant of
time. Regular black holes in which the boundary
surface is lightlike or timelike have not been found
in the literature.

\subsubsection{Solutions with boundary layers, i.e., thin shells}

Finally, it is possible and also of interest to make the transition
from an inner de Sitter core to an outer Schwarzschild,
Reissner-Nordstr\"om, or other chosen spacetime, through surface
layers, or thin shells.  Regular black holes with thin shells of
spacelike, lightlike, and timelike character have been found.

\vskip 0.3cm
\centerline{\small \it (a) Spacelike thin shells}
Following Zel'dovich's idea \cite{zel68},
Markov \cite{markov1984} suggested a
concrete upper bound for the curvature, of the
order of the Planck curvature.
After the Planck curvature bound is achieved it is
suggested that the matter turns into a de Sitter phase.
The transition is made through a spacelike thin shell.
This was developed in \cite{frolov1990,frolov1996},
see also
\cite{morgan91,balb1,balb2}.

In addition,  in \cite{lakezannias85} the fitting
of closed and open sections of de Sitter space into a Schwarzschild
solution was considered, where special care was taken in the analysis of
the intrinsic properties of the spacelike surface layer of constant
curvature joining the two spacetimes. See also \cite{burinskii2001}
for a general discussion including the Kerr-Newman metric.

\vskip 0.3cm
\centerline{\small \it (b) Lightlike thin shells}
Even before Dymnikova \cite{dy92} developed her regular black hole
with smooth features, Gonzalez-Diaz in 1981 \cite{gd81} took interest
in finding a regular black hole. He tried a solution by direct
matching of de Sitter spacetime with the Schwarzschild solution on the
horizon, a null surface.  Shen and Zhu reanalyzed later this soldering
of de Sitter spacetime with the Schwarzschild solution \cite{sz88},
while Shen and Tan in 1989 \cite{stan89} generalized Gonzalez-Diaz
idea to d dimensions.  It was later argued in \cite{daghig00} that a
Schwarzschild type matching can also be achieved within a more general
parametrization of the static metric by two different functions due to
the jump of the product $g_{tt}g_{rr}$.  However, Gr{\o}n and Soleng
\cite{gron85,gronsol89} showed that the direct matching onto
Schwarzschild at the horizon contained in \cite{gd81} is
incorrect. Poisson and Israel \cite{pisr88} reinforced once again that
no direct matching, of the type done in \cite{gd81,sz88,stan89}, is
possible, de Sitter spacetime cannot be soldered directly to an
exterior Schwarzschild vacuum, since the junction conditions would be
violated. It is necessary to put a thin shell of non-inflationary
material at a junction outside the event horizon.  In \cite{gallem001}
it was shown that the more general tentative matching proposed
in \cite{daghig00} is
also not possible (see in addition \cite{bron2001} for no-go
theorems).

Additional tries of the same type of
matching, now extending to the Reissner-Nordstr\"om spacetime, were
performed in 1985 by Shen and Zhu in \cite{sz51,sz52}. By
including charge the matching
problems occurring in a Schwarzschild matching may be avoided.
In 1991, Barrab\`es and Israel \cite{barisrl91} gave an example where
there is the possibility of joining correctly at a null surface and
gave interesting examples of a lightlike thin shell matching at the
Cauchy horizon (see also \cite{barrabes_hogan} for null matching), see
also \cite{frolov1996}.

\vskip 0.3cm
\centerline{\small \it (c)  Timelike thin shells}
In the context of regular black holes with boundary layers, timelike
matching is not found in the literature. So it is of interest to study
regular black hole solutions in such a case.  Regular black holes
either with a charged (usually magnetic) core or with a de Sitter core
are known, but with electric charge and a de Sitter core together, as
found here, seem to have not been explored.  We put the
electric charge on a thin shell at the surface of the object, in
one instance at the inner horizon, in all the other instances below
the inner horizon.

\subsection{General results on regular black holes}
\label{subsec-general}

General results on regular black holes, like those related to the
topology and causality of these solutions, were put forward by Borde
in an important development \cite{borde1994,borde1997}.  Also energy
conditions and other properties have been studied
\cite{senetal,zasla2009,zasla2010}.
The quasilocal energy of regular black holes has been analyzed
in \cite{balart2010}. Entropy and
thermodynamics of regular black holes have been studied in
\cite{myung1,myung2}.
For a general review on regular black holes,
including black holes with Gaussian sources, see \cite{ansoldi2008}.

\subsection{Connections to other works}
\label{subsec-connections}

An issue connected to regular black holes is quasiblack
holes. Quasiblack holes are objects whose boundary is as near a
horizon as one wants. For the outside they act as black holes,
although the inside properties are completely different
\cite{lezaslprop}. Based on a worked by Guilfoyle \cite{guilfoyle}
solutions of quasiblack holes with pressure, i.e., of relativistic
charged spheres as frozen stars, have been found in
\cite{lemoszanchin2010}. These solutions also contain, unexpectedly,
regular black holes, a particular branch of those is referred to
below.

There are some interesting investigations on the dynamics of
time-dependent bubbles, in which an observer in the outer region
describes the system as having a horizon and a black hole, and an
observer in the inner region, made of false vacuum, sees a de Sitter
universe \cite{blau,berezin,alberghi1999}.

Related to the inside of a black hole, it has been shown that, under
perturbations, an instability of the internal Cauchy horizon occurs,
and a spacelike or null true singularity emerges inside a charged
Reissner-Nordstr\"om black hole. This phenomenon is called mass
inflation as it shows an exponential growth of the local mass function
\cite{poisson1990}.

Black holes, and in particular charged black holes, singular or
regular, as elementary charged particles is an issue in itself that we
will delve into in another work.

\subsection{Layout of the paper}
\label{subsec-lay}

The present paper is organized as follows.  In
Sec.~\ref{sec-chargedfluids} we lay out some properties of cold
charged fluids in general relativity.  In
Sec.~\ref{sec-basicequations} the basic equations describing a charged
fluid are written. Spherically symmetric equations are then set up in
Sec.~\ref{sec-sphericalsolution}. Then in Sec.~\ref{sec-FV} regular
charged black hole solutions are presented.  In
Sec.~\ref{sec-simplifyingassumptions} we give the ans\"atze used, such
as that the matter fluid is described by a $\Lambda$ term or false
vacuum form.  Sec.~\ref{sec-QBH} is devoted to present the regular
charged black hole solutions, to the study of the main properties of
these black holes, and to comment on the relation of the present
solutions to other regular black hole solutions found in the
literature and to some of the solutions of Guilfoyle.  In
Sec.~\ref{sec-conclusion} we conclude.

\section{Charged fluids}
\label{sec-chargedfluids}

\subsection{Basic equations}
\label{sec-basicequations}
The cold charged fluids considered in the present work are described by
the Einstein-Maxwell equations with matter, which can be written as
\begin{eqnarray}
& &G_\munu=  8\pi \,\tau_\munu\, ,
\label{einst}\\
& & \nabla_\nu F^\munu = 4\pi\,J^\mu\,, \label{maxeqs}
\end{eqnarray}
where Greek indices $\mu, \nu$, etc., run from $0$ to $3$,
$0$ corresponding to a timelike coordinate $t$.
$G_\munu=R_\munu-\frac{1}{2}g_\munu R$ is the Einstein tensor, with
$R_\munu$ being the Ricci tensor, $g_\munu$ the metric, and $R$ the
Ricci scalar. We put both the speed of light $c$ and the
gravitational constant $G$ equal to unity throughout.
The tensor $\tau_\munu$ is the energy-momentum tensor
which here can be decomposed into two parts $E_\munu$
and $T_\munu$,
\begin{equation}
\tau_\munu=
\left( T_\munu+ E_\munu\right)\,.
\label{generaltmunu}
\end{equation}
$E_\munu$ is the
electromagnetic energy-momentum tensor, which can be written in the form
\begin{equation}
E_\munu= \frac{1}{4\pi}\left(
{F_\mu}^\rho F_\nu{_\rho} -\frac{1}{4}g_\munu
F_{\rho\sigma} F^{\rho\sigma}\right)\, ,\label{maxemt}
\end{equation}
where the Maxwell tensor is
\begin{equation}
F_\munu = \nabla_\mu {\cal A}_\nu -\nabla_\nu {\cal A}_\mu\, ,
\label{ddemfield}
\end{equation}
$\nabla_\mu$ representing the covariant derivative, and ${\cal A}_\mu$
the
electromagnetic gauge field. In addition,
\begin{equation}
J_\mu = \rho_{\rm e}\, U_\mu\, \label{current}
\end{equation}
is the current density, with $\rho_{\rm e}$ and $U_\mu$ being
respectively
the electric charge density and the fluid
four-velocity. $T_\munu$ is the material
energy-momentum tensor, which, for the purpose of the present work, is
taken in the form of an isotropic fluid
\begin{equation}
T_\munu = \left(\rho_{\rm m}+p\right)U_\mu U_\nu +p  g_\munu \, ,
       \label{fluidemt}
\end{equation}
where $\rho_{\rm m}$ is the fluid matter-energy density, and $p$ is
the isotropic fluid pressure.

\subsection{Spherical equations: general analysis}
\label{sec-sphericalsolution}

We particularize here the study to spherically symmetric systems,
where the charged fluid distribution is bounded by a spherical
surface $S$, whose exterior region can be described by the
electrovacuum Reissner-Nordstr\"om metric.  We then assume that the
spacetime inside $S$ is static and spherically symmetric, so that the
metric is conveniently written in a Schwarzschild-like form, namely,
\beq
ds^2 = -B(r)\,dt^2 + A(r)\, dr^2 + r^2\,
(d\theta^2+\sin^2 d\phi^2)\, ,
\label{metricsph}
\eeq
where $r$ is the usual radial coordinate, $A$ and $B$ are functions which
depend upon $r$ only, and $d\theta^2+\sin^2 d\phi^2$
is the metric of the unit sphere, with $\theta$ and $\phi$ being
the spherical angles.
The gauge
field ${\cal A}_\mu$
assumes the form
\beq
{\cal A}_\mu = -\phi(r)\,\delta_\mu^t\, ,\label{gauge2}
\eeq
where $\phi(r)$ is the electric potential, and here $\delta$ is
the Kronecker delta.
The four-velocity $U_\mu$ in turn is
\beq
U_\mu =  -\sqrt{B(r)\,}\, \delta_\mu ^t\, .\label{veloc2}
\eeq

Define $r_0$ as the radius of the spherical surface $S$. Consider
the region $r< r_0$. This region is filled with matter, i.e., filled
with 
a
static charged perfect fluid distribution with spherical symmetry. Then,
the relevant Einstein-Maxwell field equations for the metric
\eqref{metricsph} can be
found. To find the set of basic equations we note that the $tt$ and $rr$
components of Einstein equation \eqref{einst} furnish the following
relations
\beqa
&&\hspace*{-1.3cm}\frac{B^\prime(r)}{B(r)}+\frac{A^\prime(r)}{A(r)} =
8\pi
r\,A(r)
\Big[\rho_{\rm m}(r) +p(r)\Big]\, ,\label{einsteq1}\\
&&\hspace*{-1.3cm}
  \Big(r\, A^{-1}(r)\Big)^\prime = 1- 8\pi\, r^2\left(
\rho_{\rm m}(r) + \frac{Q^2(r) }{8\pi r^4}\right)\, , \label{einsteq2}
\eeqa
where a prime denotes the derivative with respect to the radial
coordinate $r$.
The only nonzero component of the Maxwell equation
\eqref{maxeqs} furnishes
\beq
Q(r) =  4\pi\int_0^r { \rho_{\rm e}(r)\sqrt{A(r)} \,r^2 dr}=
  \frac{r^{2}\,\phi^\prime (r)}{\sqrt{B(r) \,A(r)}}\, ,
\label{chargedef}
\eeq
where an integration constant was set to zero. $Q(r)$ is
the total electric charge inside the surface of radius $r$.
The conservation equations $\nabla_\nu T^{\mu\nu} =0$, together with
the Maxwell equation, give
\beq
  2p^\prime(r) + \frac{B^\prime(r)}{B(r)} \Big [\rho_{\rm m}(r) +
p(r)\Big] - 2\frac{\phi^\prime(r) \rho_{\rm e}(r)} {\sqrt{B(r)}}=0,
        \label{conserv1}
\eeq
which is the only non-identically zero component of the conservation
equations. The other nonzero component of the
Einstein-Maxwell equations give
an additional relation which is not independent of the above set of
equations.
Hence, considering that Eq.~\eqref{chargedef} gives $\phi(r)$ in terms
of 
$
\rho_{\rm e}(r)$, we are left with a system of three
differential equations, Eqs.~\eqref{einsteq1}, \eqref{einsteq2} and
\eqref{conserv1},
for the five unknowns $A(r)$, $B(r)$, $\rho_{\rm m}(r)$, $p(r)$ and
$\rho_{\rm e}(r)$. Below we study a particular choice of relations to fix
the two degrees of freedom and show that with an appropriately chosen
equation of state $p=f(\rho_{\rm m})$ and further assumptions on the
functions $\rho_{\rm m}(r)$ and $\rho_{\rm e}(r)$, one can find regular
black hole solutions.
For $r\geq r_0$, we assume electrovacuum. Then the metric and
the electric potential for $r>r_0$,
are given by the Reissner-Nordstr\"om solution
$
B(r)={A^{-1}(r)} = 1 -\dfrac{2m}{r}+\dfrac{q^2}{r^{2}}$,
$\phi(r)= \dfrac{q}{r}+\phi_0
$,
$\phi_0$ being an arbitrary constant which defines the zero of the
electric potential, and that, in the exterior Reissner-Nordstr\"om
region of the spacetimes as we consider here, can be set to zero.
The parameters $m$ and $q$ are the mass and charge of
the Reissner-Nordstr\"om solution, respectively.
The matching of the interior and exterior solutions
at $r_0$ will have then to be made.

\section{Electrified false vacuum solutions: Regular charged
black holes}
\label{sec-FV}

\subsection{Simplifying assumptions}
\label{sec-simplifyingassumptions}

In order for the energy-momentum tensor to satisfy the false vacuum 
condition
the energy density and pressure of the
the fluid~\eqref{fluidemt} must obey the following relation
\beqa
  & & \rho_{\rm m}(r) + p(r)=0 \, ,
\label{falsevac2}
\eeqa
valid for all $r$. Indeed, in the true vacuum, for $r\geq r_0$, the
condition is trivially obeyed, since $\rho_{\rm m}(r)=0$ and $p(r)=0$.
In the matter, $r<r_0$, the above equation, Eq.~\eqref{falsevac2}, is
equivalent to a $\Lambda$ term, and can be interpreted as representing
a false vacuum condition. Indeed, a false vacuum is usually given in
the form $T_t^t = T_r^r$ and $T_\theta^\theta= T_\phi^\phi$, the same
as Eq.~\eqref{falsevac2}.  The ansatz~\eqref{falsevac2} can be
interpreted as an equation of state, and supplies a constraint among
the five unknown quantities of the problem.

However, since we have just three equations relating the five
unknowns, we need another input. Such a degree of freedom is related
to the charge density and in order to close the system usually the
function $\rho_{\rm e}(r)$, or $Q(r)$ if one prefers, is furnished by
making an additional hypothesis.  In the present case, one of the
simplest ansatz one can make is through the following equation
\beq
8\pi \rho_{\rm m}(r) + \frac{Q^2(r)}{ r^4} =
\frac{3}{R^2} \, , \label{constantenergy}
\eeq
for $r\leq r_0$, and
where $R$ is a constant to be determined. This resembles the
Schwarzschild assumption of constant energy density
for the first interior solution found within general
relativity (see, e.g., \cite{stephanietal}). In the present
case, the energy density includes not only the energy density due to
the mass distribution, but it also bears the electromagnetic
energy-density carried by the electric field and represented by the
term $\dfrac{Q^2(r)}{8\pi\,r^4}$.
Now, after the assumption \eqref{falsevac2},
Eqs.~\eqref{conserv1} and \eqref{chargedef} give
\beq
p^\prime(r)= \frac{\phi^\prime(r) \rho_{\rm e}(r)}
{\sqrt{B(r)}}= \frac{QQ^\prime}{4\pi r^4}.  \label{conserv2}
\eeq
This means that the pressure gradient $p'(r)$ is proportional to the
charge density.
Finally, with condition \eqref{falsevac2}, it is obtained from
Eq.~\eqref{einsteq1} that the metric
coefficients are related by $A(r)\, B(r)= {\rm constant}$. This
constant can be
set to unity by a re-parametrization of the coordinate $t$. Then,
\beq
B(r)= A^{ -1}(r), \label{coeffB}
\eeq
and we can now complete the process of integration of the
full set of equations.

\subsection{Electrified de Sitter false vacuum solutions: Regular charged
black holes}
\label{sec-QBH}

\subsubsection{Finding the solutions }
\label{sec-fts}
Bearing in mind that for $r> r_0$ it is
a true electrovacuum solution and thus
the Reissner-Nordstr\"om solution,  and using the
ans\"atze \eqref{falsevac2} and \eqref{constantenergy},
equations~\eqref{einsteq1} and \eqref{einsteq2} integrate to
\beq
B(r) = A^{-1}(r)  = \left\{\begin{array}{l l}
\displaystyle{ 1-\frac{r^2}{R^2}}, &  {r\leq r_0},\vspace{.13cm}\\
\displaystyle{1- \frac{2m}{r} + \frac{q^2}{r^2}}, & \  {r\geq r_0},
\end{array}
\right.
\label{BAsol1}
\eeq
where an integration constant was set to zero to avoid a spacetime
singularity at $r=0$. By joining the interior and the exterior metrics at
$r=r_0$, i.e., by equating the metric coefficients (functions $B(r)$ and
$A(r)$, see Appendix A) in Eq.~\eqref{BAsol1} with the coefficients of
the
exterior Reissner-Nordstr\"om metric,
the constant $R$ in Eq.~\eqref{BAsol1} is found in terms of
the parameters $r_0$, $m$, and $q$, through the equation
\beq
\frac{1}{R^2} = \frac{1}{r_0^3}\left( 2m - \frac{q^2}{r_0}\right)\, .
\label{Rm2}
\eeq
From the above set of equations we can infer several things. From
Eq.~(\ref{BAsol1}) one can infer that
\beq
r_0\leq R\, ,
\label{less}
\eeq
otherwise one would have three zeros, i.e., three horizons, for the
metric corresponding to (\ref{BAsol1}), an odd number of
horizons, and this is not possible
for regular black holes  (as well as for the electrovacuum ones
in general relativity) in asymptotically flat
spacetimes. For $r_0\leq R$ one can also infer from Eq.~(\ref{Rm2}) that
$1- \frac{2m}{r_0} +
\frac{q^2}{r_0^2}\geq 0$ which means that either $r_0$ is
inside  or on the inner, or Cauchy, horizon $r_-$, or
$r_0$ is
outside or on the outer, event, horizon $r_+$. Since we
want black holes we search for solutions in which
$r_0\leq r_-$, but the procedure dictates what kind of solutions there 
are.

Our choices (\ref{falsevac2}) and (\ref{constantenergy})
for $r\leq r_0$ can
now be written for $\rho_{\rm m}(r)$ and $p(r)$ in terms of the electric
charge $Q(r)$, namely,
\beqa
8\pi \rho_{\rm m}(r) = \frac{3}{R^2} -\frac{Q^2(r)}{r^4},
\label{rhom1} \\
8\pi p(r) =  -\frac{3}{R^2} + \frac{Q^2(r)}{r^4}.
\label{pressure1}
\eeqa
Now using Eqs.~\eqref{rhom1}-\eqref{pressure1}, as well as
Eq.~\eqref{conserv2}, we can determine $Q(r)$. The result is $Q(r)=0$ in
the
region where there is matter, where $p\neq 0$,
$r<r_0$. And
$Q'(r)=0$ in the
electrovacuum region, where $p=0$, $r\geq r_0$.
With this, one can build a solution where $Q(r)$ is a step function, more
precisely, it may be represented by the step Heaviside theta function
\beq
Q(r) =  \left\{\begin{array}{l l}
\displaystyle{0}, &  {r < r_0},\vspace{.13cm}\\
\displaystyle{q }, & \  {r\geq
r_0}\,.
\label{chargefunction1}
\end{array}
\right.
\eeq
In such a case, the charge density is proportional to a Dirac
delta function, the charge $\rho_{\rm e}(r)$
is spread at the surface $r=r_0$,
%i.e., $ \rho_{\rm e}(r) = \dfrac{q}{4\pi r_0^2
%\sqrt{A(r_0)}}\,\delta(r-r_0)\, $ whose volume integration, see
%Eq.~\eqref{chargedef}, gives exactly the result shown in
%Eq.~\eqref{chargefunction1}.
Furthermore, one gets $p'(r)=0$ everywhere
throughout the spacetime, except
at $r=r_0$ where it is the product of the Heaviside theta, or
step, function by the Dirac delta function.
%$Q(r) \sim \theta(r-r_0)]$ by the Dirac delta function [$ Q'(r) \sim
%\rho_{\rm e}(r)\sim \delta(r-r_0)$].

The electrical potential $\phi(r)$ is obtained from
Eqs.~\eqref{chargedef} and \eqref{chargefunction1}, and may be written
as
\beq
\phi(r) =\left\{\begin{array}{l l}
\displaystyle{ -\frac{q}{r_0}} , &  {r\leq r_0},\vspace{.13cm}\\
\displaystyle{-\frac{q}{r} }, & \  {r\geq r_0} \, .
\end{array}
\right.
\eeq
As it is known, the exterior geometry, i.e., for $r>r_0$, depends on
$\phi^\prime(r_0)$ only through the total charge $q=Q(r_0) $, i.e., in
that region, the only important quantity is the value of the electric
field on the surface $r=r_0$, as expected.

Now, we present other important relations between
the parameters. At $r_0$,
Eqs.~(\ref{rhom1})-(\ref{pressure1}) turn into
\beqa
8\pi \rho_{\rm m}(r_0) = \frac{3}{R^2} -\frac{q^2}{r_0^4},
\label{rhom2} \\
8\pi p(r_0) =  -\frac{3}{R^2} + \frac{q^2}{r_0^4}.
\label{pressure2}
\eeqa
Using the appropriate soldering conditions \cite{israel} one should have
$p(r_0)=0$ at $r_0$ (see also Appendix A). From Eq.~(\ref{pressure2})
this means
\beq
R\,q=\sqrt3\,r_0^2\,,
\label{qtor0ratio}
\eeq
where we assumed $q>0$ without loss of generality.
This also means from Eq.~(\ref{rhom2}) that $\rho_{\rm
m}(r_0)=0$. Using Eq.~(\ref{qtor0ratio}) in Eq.~(\ref{Rm2}) we
find
\beq
m\,r_0=\frac{2}{3}\,q^2\,.
\label{mtor0ratio}
\eeq

Interestingly, in the case $R=r_0$, putting the results of
Eqs.~\eqref{qtor0ratio} and \eqref{mtor0ratio} into the horizon
equation (i.e., the radii for which $A(r)=0$), the boundary surface
$r=r_0$ coincides with the inner horizon, i.e., the Cauchy horizon
$r_-$.  For all the other cases $r_0$ is inside $r_-$.  The event
horizon $r_+$, if there is one, is then always outside the matter.
One can also check that the second fundamental form is continuous at
$r_0$. This is equivalent to $B^\prime(r)$ being continuous at $r_-$
(see Appendix \ref{appA}).  It is easy to verify that this is true for
the values given in Eqs.~(\ref{qtor0ratio})-(\ref{mtor0ratio}) and
only for these values. In fact, as we show in Appendix \ref{appA}, the
condition of zero pressure at the surface $r=r_0$ implies that the
metric fields $g_{tt}(r)=B(r)$ and $g_{rr}=1/B(r)$ are $C^1$ functions
at the boundary, which means that the Israel junction conditions
\cite{israel} are satisfied even in the case $r_0$ is a lightlike
surface.

Note also that when there is no charge there is no black hole since
for $q=0$ one has $r_0=0$ and $m=0$, leaving a Minkowski vacuum
spacetime. Thus the limit of zero charge is not a regular
Schwarzschild black hole, but rather a Minkowski spacetime.  This is
in tune with the claims of Gr{\o}n and Soleng \cite{gron85,gronsol89}
and Poisson and Israel \cite{pisr88} that one cannot have pure de
Sitter joined at the event horizon $r_+$ to a Schwarzschild vacuum
spacetime.  However, as we show here, one can have pure de Sitter plus
a charged shell joined to Reissner-Nordstr\"om vacuum spacetime.

\subsubsection{The final solution in brief}
\label{sec-tfsib}
There are four parameters: $m$, $q$, $r_0$ and $R$.
From Eqs.~(\ref{qtor0ratio})-(\ref{mtor0ratio})
the de Sitter radius $R$ and the black hole
mass $m$ are fixed respectively by
\beq
R=\sqrt3\,r_0^2\,\frac{1}{q}\,,
\label{qtor0ratio2}
\eeq
assuming positive $q$, and
\beq
m=\frac{2}{3\,r_0}\,q^2\,.
\label{mtor0ratio2}
\eeq
So there are two free parameters, $r_0$ and $q$ say.
Fixing
$q$ one can change $r_0$. From Eq.~(\ref{less}) and
Eq.~(\ref{qtor0ratio}) one finds
\beq
r_0\geq \frac{\sqrt{3}}{3}{q}\,,
\label{radmin}
\eeq
as well as,
\beq
m\leq \frac{2}{\sqrt{3}}{q}\,.
\label{massmin}
\eeq

To summarize the solution we display it here wholly.
Defining, as usual, the Heaviside $\theta$ function as
$\theta(r-r_0)=1$ when $r-r_0\geq 0$ and $\theta(r-r_0)=0$ when $r-r_0< 
0$,
and the Dirac delta function as the derivative of it,
$\delta(r-r_0)=[\theta(r-r_0)]^{\,\prime}$, where
a $^{\,\prime}$ denotes radial derivative,
one can write all the functions in succinct form.
The line element is
\beq
ds^2 = -B(r)\, dt^2 +  B^{-1}(r) \, dr^2 + r^2 (d\theta^2+\sin^2
d\phi^2) 
\,,
\label{metricsphfinal}
\eeq
where
\beqa
B(r) = & &
\left(1-\frac{r^2}{R^2}\right)\theta(r_0-r)\nonumber\\
& &+\left(1- \frac{2m}{r} +
\frac{q^2}{r^2}\right)\left[1-\theta(r_0-r)\right]\,,
\label{BAsol1-2}
\eeqa
where $R$ and $m$ should be thought as being given in terms of $r_0$ and 
$q$
by Eqs.~(\ref{qtor0ratio2})-(\ref{mtor0ratio2}).
The electric potential is
\beq
\phi(r) =  -\frac{q}{r_0}\,\theta(r_0-r) -
\frac{q}{r}\left[1-\theta(r_0-r)\right]\,.
\label{pot1}
\eeq
\begin{figure}
\centerline{\includegraphics[scale=1.2]{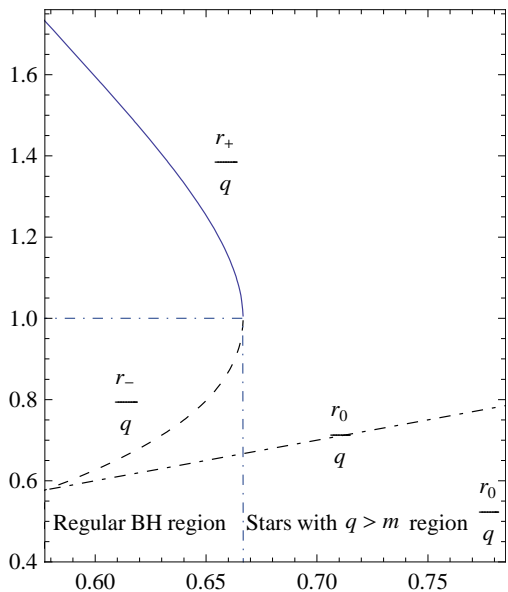}}
\caption{A plot of the several radii as a function of $\alpha=
r_0/q$. The minimum value of $\alpha$ is $\sqrt{3}/3$,
and the vertical dashed-dotted line is at $\alpha=2/3$.}
\label{plot}
\end{figure}
Also,
\beq
Q(r) =  q\,\theta(r-r_0) \,,
\label{chargefunction2}
\eeq
\beq
Q^{\,\prime}(r) =  q\,\delta(r-r_0) \,,
\label{chargeprimefunction2}
\eeq
\beq
\rho_{\rm e}(r) =
\dfrac{q\,\sqrt{B(r_0)}}{4\pi r_0^2 }\,\delta(r-r_0)\,,
\label{rhoe}
\eeq
\beq
\rho_{\rm m}(r) = \frac{q^2}{8\pi\,r_0^4} \left(1 -
\theta^2(r-r_0)\right)
\theta(r_0-r)\,,
\label{rhom02}
\eeq
\beq
p(r) = -\frac{q^2}{8\pi\,r_0^4} \left(1 -
\theta^2(r-r_0)\right)
\theta(r_0-r)\,,
\label{pressure02}
\eeq
\beq
p^{\,\prime}(r)=\frac{q^2}{4\,\pi\,r_0^4}\,\delta(r-r_0)\,,
\label{pressureprime}
\eeq
see Appendix \ref{appB} for how to obtain Eq.~(\ref{pressureprime}) from
Eq.~\eqref{pressure02}.
Using the fact that $\theta^2(r-r_0)=\theta(r-r_0)$ (this
can only be used if derivatives are not to
be taken), Eqs.~(\ref{rhom02})-(\ref{pressure02}) simplify to
$8\pi\,\rho_{\rm m}(r) = -8\pi\,p(r)=
\displaystyle{\frac{3}{R^2}}\left[1-\theta(r-r_0)\right].$

For a range of parameters,
the solution represents regular black holes with a de Sitter core, and
an electric energyless matter coat at $r_0$, and Reissner-Nordstr\"om all
the way up.  When there is no charge there is no regular black hole
since for $q=0$ one has $r_0=0$ and $m=0$, leaving a Minkowski vacuum
spacetime.

\subsubsection{The solutions and their properties}
\label{sec-pots}

Since we are not interested in the case $q=0$,
as it gives the trivial Minkowski solution, we can parametrize all
quantities in terms of $q$. Eq.~(\ref{radmin}) tell us that $r_0$
obeys, $r_0\geq \frac{q}{\sqrt3}$.
Thus let us put
\beq
r_0={\alpha}\, q\,,
\label{alpha}
\eeq
for some ${\alpha}$, with ${\alpha}\geq1/\sqrt3$.
Then, Eq.~(\ref{qtor0ratio2}) gives
\beq
R={\sqrt3}{\alpha}^2\, q\,,
\label{morealpha2}
\eeq
and Eq.~(\ref{mtor0ratio2}) furnishes
\beq
m=\frac{2}{3{\alpha}}\, q\,.
\label{morealpha1}
\eeq
Now, we have to compare $r_0$ with $r_-$ and $r_+$.
These latter radii are given by
\beq
r_\pm=m\pm\sqrt{m^2-q^2}\,.
\label{hor}
\eeq
Putting Eq.~(\ref{morealpha1}) into Eq.~(\ref{hor}) yields
$r_-$ as
\beq
r_-=\left(
\frac{2}{3{\alpha}}-
\sqrt{\frac{4}{9{\alpha}^2}-1}\right)\,q\,,
\label{hor2-}
\eeq
and $r_+$ as
\beq
r_+=\left(
\frac{2}{3{\alpha}}+
\sqrt{\frac{4}{9{\alpha}^2}-1}\right)\,q\,.
\label{hor2+}
\eeq
A plot of the several radii as a function of $\alpha=
r_0/q$ is given in Fig.~\ref{plot}.

\begin{figure}
\centerline{\includegraphics[scale=1.2]{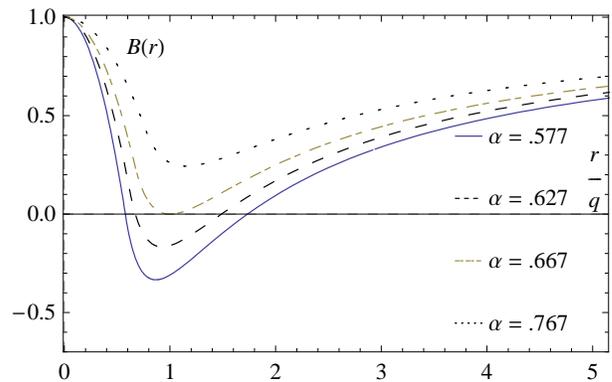}}
\caption{A plot of $B(r)$ for four
different $\alpha\equiv \frac{r_0}{q}$: $\alpha
=\sqrt{3}/3$ (solid line), $\alpha= \sqrt{3}/3+ .1$ (dashed line),
$\alpha = 2/3$ (dash-dotted line), and $\alpha = 2/3+.1$ (dotted line). }
\label{plotB}
\end{figure}

The solutions can be divided into four classes, which are listed and 
briefly discussed in the following. Figure \ref{plotB} plots the function
$B(r)$ and Figure \ref{penr} draws the Carter-Penrose diagrams for each
class. Figures \ref{plot}-\ref{penr} help in the explanation of the
classes.

\vskip 0.3cm
\centerline{\it (a) Regular nonextremal black holes}
\centerline{\it with a
lightlike matter boundary at the inner horizon}
\noindent
When ${\alpha}=r_0/q$
has its minimum value $r_0/q=\sqrt3/3$, then the radius $r_0$ of the 
matter coincides with the Cauchy horizon $r_-$, so that $r_-/q=\sqrt3/3$.
There is matter up to $r_-$. The event horizon at $r_+$ is at a larger
radius. From Eq.~(\ref{hor2+}) with ${\alpha}=\sqrt3/3$, one finds
$r_+/q=3r_-/q=\sqrt3$. This is a solution for a perfectly regular black
hole, Reissner-Nordstr\"om outside the matter. One can also find a
relation between the surface charge density and the horizon radii. From
Eq.~(\ref{rhoe}) defining a proper surface electric charge density
as $\rho_{\rm e}(r) = \sigma_{\rm e}\,\sqrt{B(r)}\,\delta(r-r_0)$,
one finds $\sigma_{\rm e}=\frac{q}{4\pi r_0^2}=
\frac{1}{4\pi}\frac{3}{q}$. This can also be put in the
form $\sigma_{\rm e}=\frac{1}{4\pi}\frac{\sqrt3}{r_-}$.
Using $r_+/q=3r_-/q=\sqrt3$ and squaring the result, one obtains
$\sigma_{\rm e}^2=\frac{1}{16\pi^2}\frac{r_+}{r_-^3}$.
For this solution, in which $R^2 = q^2/3$, one can show that the surface
gravities of the de Sitter horizon and of the Cauchy Reissner-Nordstr\"om
horizon are equal. The surface pressure and surface energy density are
both zero at $r_-$.

\vskip 0.3cm
\centerline{\it (b) Regular nonextremal black holes}
\centerline{\it with a timelike matter boundary}
\centerline{\it inside the inner horizon}
\noindent
For larger $r_0/q$, i.e., larger $r_0$, the Cauchy horizon $r_-$ grows 
faster and remains outside the radius of the matter, $r_0<r_-$. There is
matter up to $r_0$ and then outside the matter stand two horizons.
This is also a solution for a perfectly regular black hole,
Reissner-Nordstr\"om outside the matter.

\begin{widetext}
\vspace*{-.3cm}
\begin{figure}[htb]
\begin{center}
\includegraphics[scale=.43]{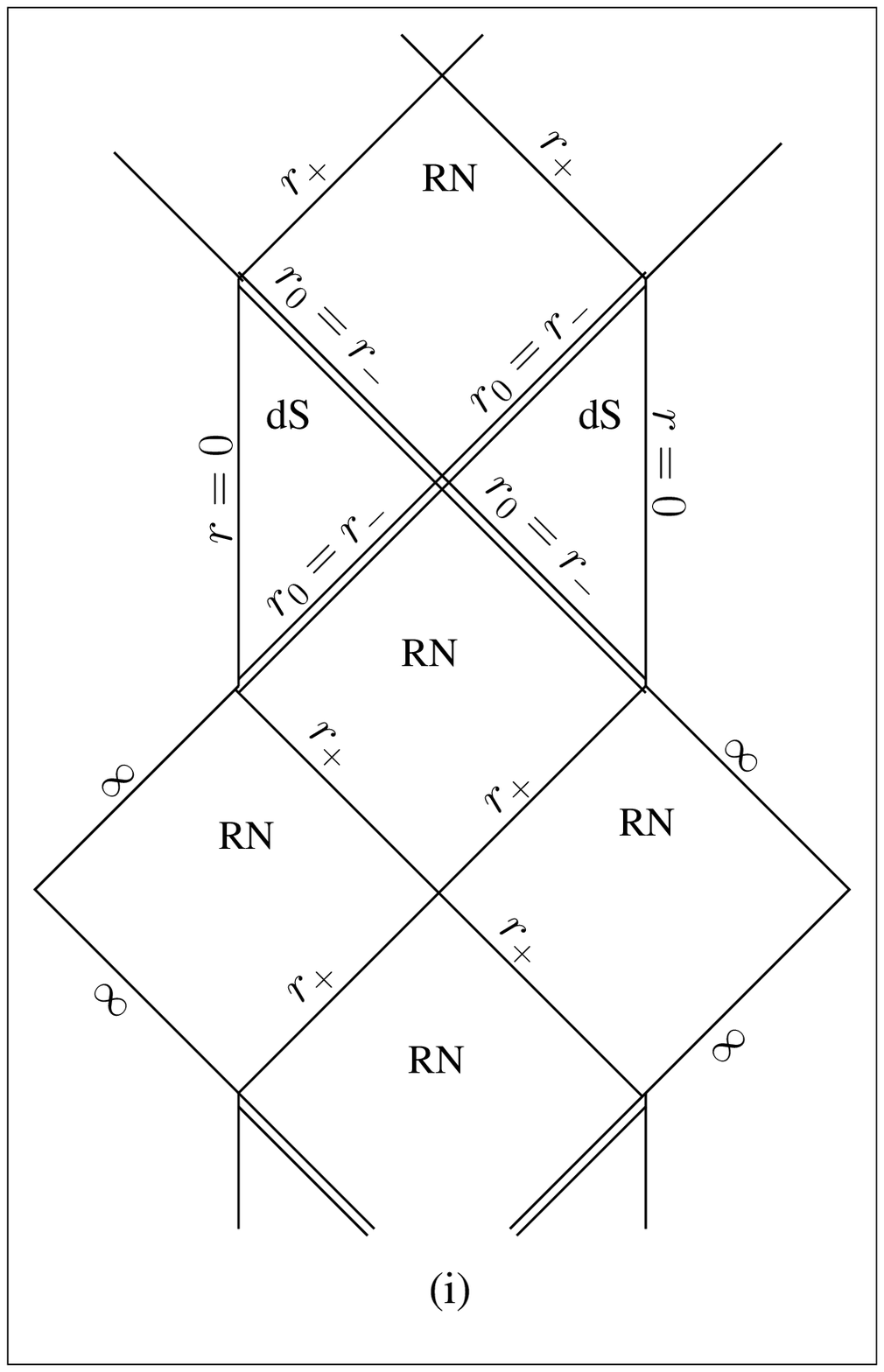} \hskip 1.3cm
\includegraphics[scale=.43]{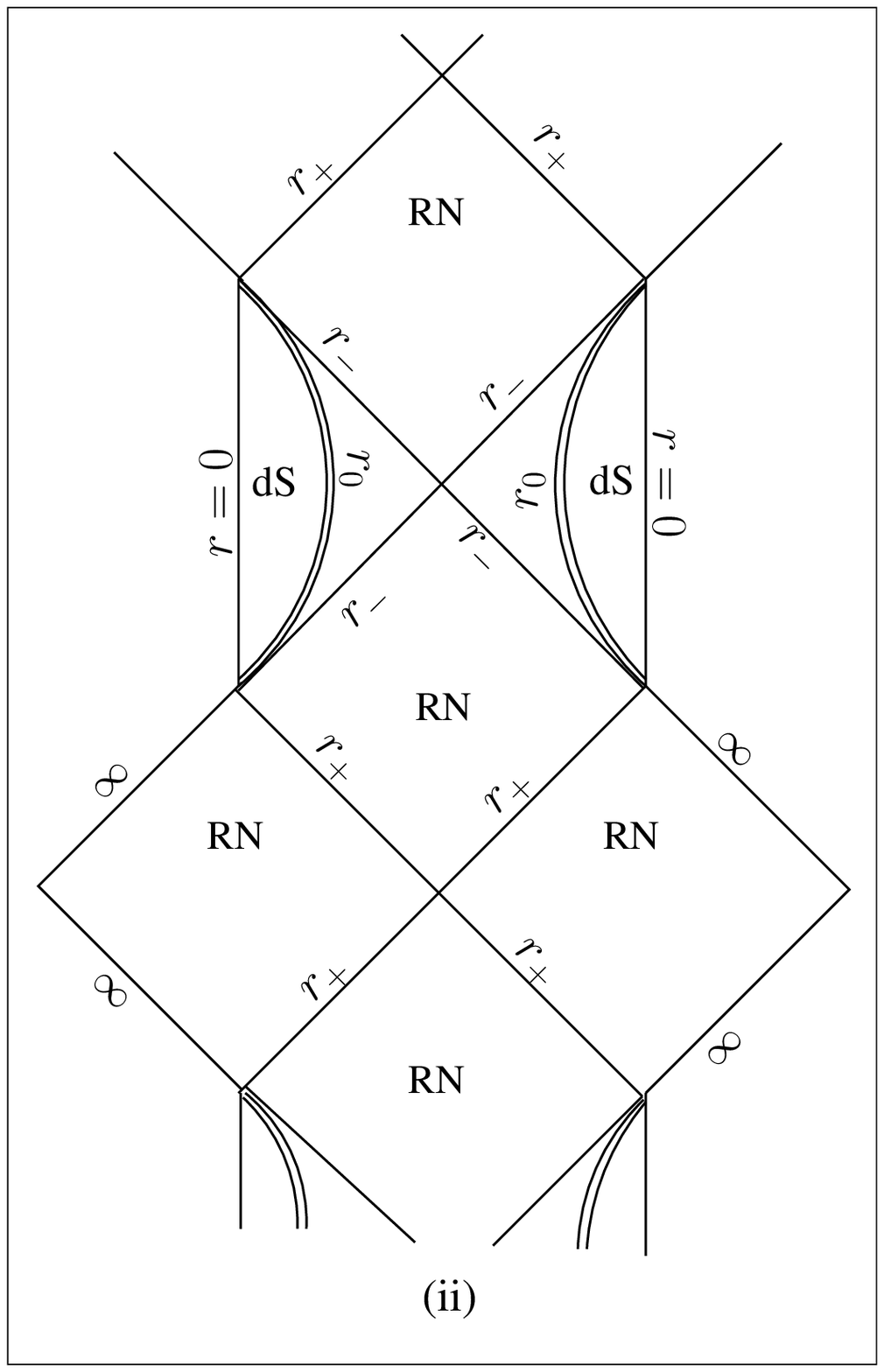}\\
\vskip .2cm
\hskip -.35cm
\includegraphics[scale=.44]{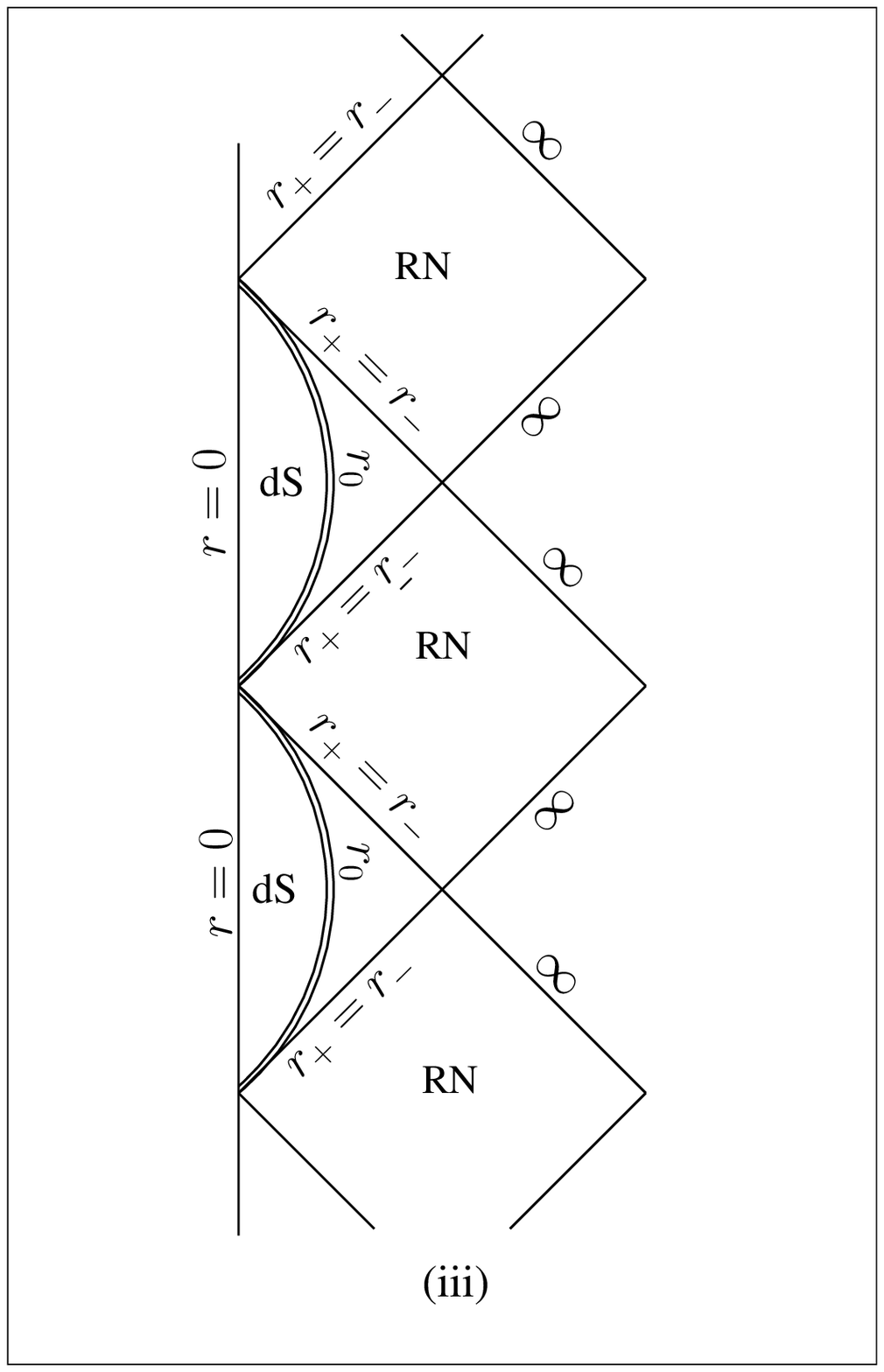} \hskip1.7cm
\includegraphics[scale=.44]{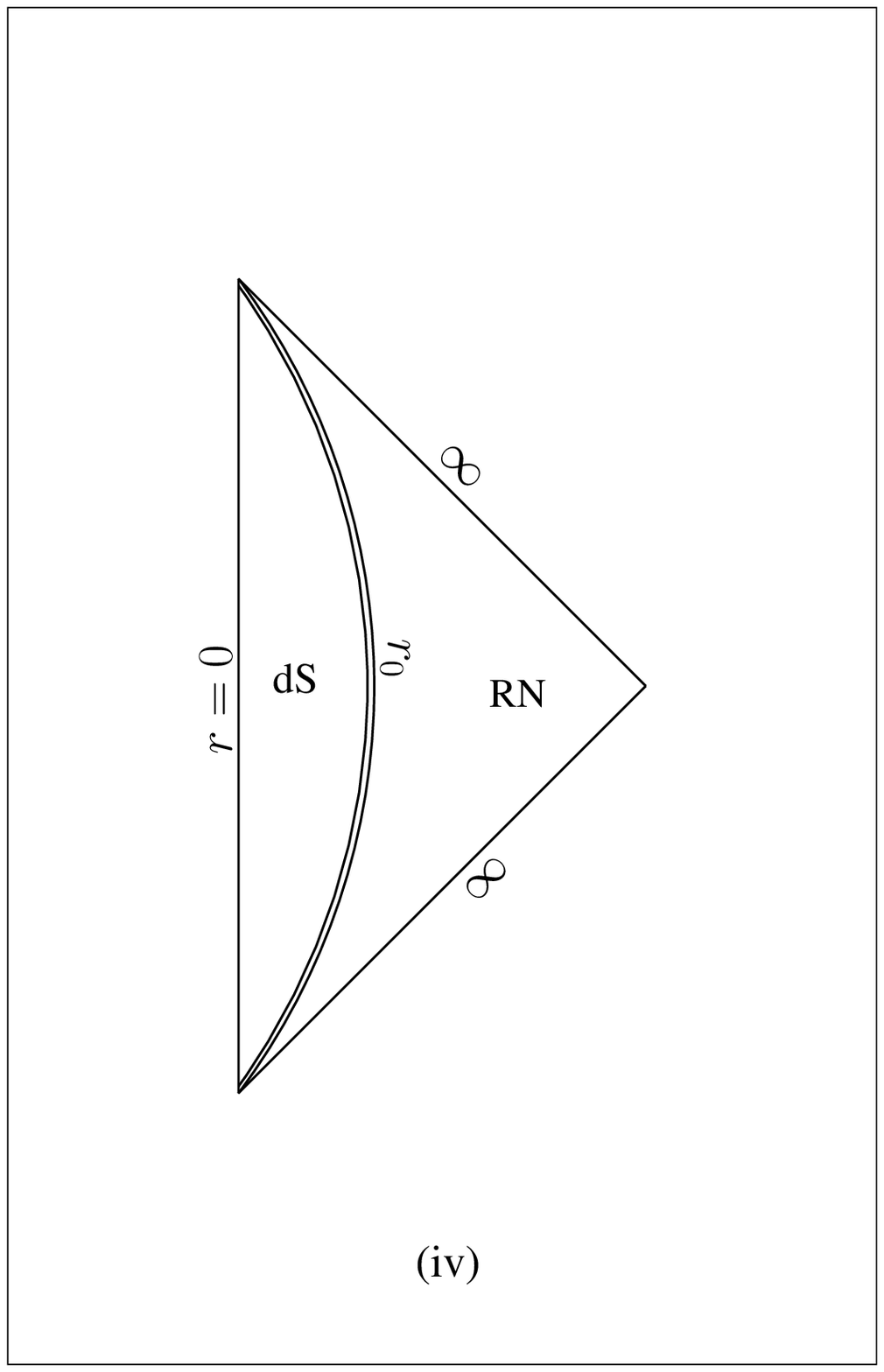}
\caption{The Carter-Penrose diagram in the possible four distinct
situations, three for regular black holes and one for a star: (i) The
regular nonextremal black hole with a lightlike matter boundary, where
$r_0=r_-$. (ii) The regular nonextremal black hole with a timelike
matter boundary, where $r_0<r_-$. (iii) The regular extremal black
hole with a timelike matter boundary, where $r_0<r_-=r_+$. (iv) The
regular overcharged star with a timelike matter boundary, where there
is no $r_\pm$. In all cases, $r_0$ represents the surface matching an
interior de Sitter spacetime region to an exterior
Reissner-Nordstr\"om region. Note the regular black hole constraint
$r_0\leq r_-$. }
\label{penr}
\end{center}
\end{figure}

\end{widetext}

\centerline{}
\centerline{}
\vskip 0.3cm
\centerline{\it (c) Regular extremal black holes}
\centerline{\it with a timelike matter boundary}
\centerline{\it inside the double horizon}
\noindent
When $r_0/q=2/3$ then the Cauchy horizon and the event horizon
coincide, the solution represents an extremal regular black
hole. Since $r_-=r_+=m=q$ we have $r_0=\frac23 r_+$. The horizon is
now at the furthest coordinate distance from the surface of the matter
at $r_0$. In fact the horizon is at an infinite proper distance from
the surface of the matter.

\vskip 0.3cm
\centerline{\it (d) Regular overcharged stars}
\centerline{\it with a timelike matter boundary}
\noindent
For $r_0/q>2/3$ one gets from Eq.~(\ref{mtor0ratio2})
that $q>m$, the solution has charge greater than mass.
Thus from Eq.~(\ref{hor}) there are no horizons
and so no black holes. An overcharged star, with no horizons, pops up.
The star that was hidden behind a horizon comes into light.
The horizons have disappeared.
It is of remark that $r_0$ of the first star is still
smaller than $r_-$ of the extremal regular black hole.
Perhaps this is no surprise as we are familiar with the
fact that the Reissner-Nordstr\"om solution has the same
feature, after the extremal horizon the singularity
at $r=0$ becomes bare for the overcharged solutions.

\vskip 0.3cm
Several other features should be mentioned
and emphasized. \hfill

\noindent
(i) For a range of parameters the solutions are thus regular
electrically charged black hole solutions. They are built from false
vacuum up to, but not at, $r_0$.  The metric for $r<r_0$ is the de
Sitter metric, where the isotropic pressure is constant
($p(r)=-\rho_{\rm m}(r) = 3/8\pi R^2$), and goes to zero at
$r_0$. Furthermore, since the charge density $\rho_{\rm e}(r)$ is a
Dirac delta function centered in $r=r_0$, the total charge $q$ is
distributed uniformly on the surface $r=r_0$.  At $r_0$ there is thus
a thin electrical layer of an energyless field, and exterior to it it
is pure Reissner-Nordstr\"om, with two horizons at $r_-$ and $r_+$.
The radius to charge ratio and mass to charge ratio of the solution
are well defined quantities.

\noindent
(ii)  From Eq.~(\ref{morealpha1}) we see that as
$r_0$ increases the mass $m$ decreases. This is due
to the fact that the pressure, and thus the density,
both decrease fast as $r_0$ increases.

\noindent
(iii) The limit
of zero charge of these solutions
is a Minkowski spacetime, rather than
a Schwarzschild spacetime.

\noindent
(iv) These regular charged black hole solutions have
boundaries which are either timelike or, in
one instance, lightlike. The boundaries for
regular black holes found in the literature
are spacelike.

\noindent
(v) Note that, if the charge $q$ is the elementary charge,
i.e., the electron charge $e$,
then Eq.~(\ref{radmin}) gives that
the radius $r_0$ of the particle is of the order of the Planck
radius and from Eq.~(\ref{massmin}) the mass $m$ is of the order of the
Planck mass. The solution could then be a model for a
heavy elementary charged particle.

\noindent
(vi) Due to the fact that the matter of the regular
black hole is in the inside  or at the Cauchy horizon,
these solutions may suffer  the mass inflation instability
\cite{poisson1990}.

\noindent
\subsubsection{Comments on the regular nonextremal black hole with a
lightlike matter boundary at the inner horizon $r_-$}
\label{sec-conn}

Since the regular nonextremal black hole with a lightlike matter boundary
at the inner horizon $r_-$  (see part {\it (a)} of Section
\ref{sec-pots}) has some history we comment on it. In {\it (a)} below we
review and connect our solution to other works that also found this
particular solution.
In {\it (b)} we comment that this particular solution belongs
to the Weyl-Guilfoyle class of solutions.

\vskip 0.3cm
\centerline{\it (a) Other works that also found this particular solution}

As noticed in \cite{gron85,gronsol89,pisr88,gallem001,bron2001}
a direct soldering of
a matter solution onto Schwarzschild through the event horizon is not
possible. Indeed, a simple calculation \cite{pisr88} shows that the
radial pressure $p_r$ for a static spherically symmetric junction at
$r=r_0$ is $p_r=-\rho_{\rm m} \theta(r_0-r)+\frac12 \rho_{\rm m} r_0
\delta(r-r_0)$. In terms of proper radial distance $s$, with
$ds=\sqrt{A}\,dr$ (where $A$ is the metric function
defined in (\ref{metricsph})), the delta-function is
$\delta(r-r_0)={\sqrt{A}}\,\delta(s)$. At the Schwarzschild
event horizon $1/A(r_+)=0$, and thus the delta-function itself is a
singular distribution. Thus, besides being a distribution, the surface
pressure becomes singular when an event horizon is chosen as the
boundary. Therefore, the papers \cite{gd81,sz88,stan89,daghig00} must be
incorrect. A Schwarzschild event horizon is then not the place to make a
continuation to a de Sitter phase, whereas other places might give a
suitable continuation. However, when one introduces charge, and the
matter is joined instead to a Reissner-Nordstr\"om spacetime, the
impossibility of matching at the horizon can be avoided by making
$\rho_{\rm m}=0$ and $p_r=0$ at the event horizon, as well as at the
Cauchy horizon.

\noindent
(i) The two papers by Shen and Zhu of 1985
on regular black holes with Reissner-Nordstr\"om
asymptotics \cite{sz51,sz52} indeed contain, in a hidden manner, the
regular nonextremal black hole with a lightlike matter boundary at the
inner horizon $r_-$.
Both papers \cite{sz51,sz52} have the same content, but oddly the
authors do not self-cite neither paper.
The paper \cite{sz52} is more detailed.

Shen and Zhu papers \cite{sz51,sz52} give a generalization of the
Gonzalez-Diaz \cite{gd81} work (see also \cite{sz88,stan89}), by
considering the case with charge. In this way they avoid the matching
problem. In \cite{sz51,sz52} it is claimed that by taking the limit of
zero charge they recover the Gonzalez-Diaz \cite{gd81} solution. This
claim seems incorrect since Eq.~(34) of \cite{sz52} shows that in this
limit there is a singularity at $r=0$, and so there is no regular
Schwarzschild black hole.

Now, in a particular instance, the solution found in
\cite{sz51,sz52} reduces to the regular nonextremal black hole with a
lightlike matter boundary at the inner horizon $r_-$
found here.  Let us see how. 
They consider two electrical coats, one at
$r_+$, the other at $r_-$, such
that $\rho_{1 \rm e}=\sigma_{1 \rm e} \delta(r-r_+)$ and
$\rho_{2 \rm e}=\sigma_{2 \rm e}
\delta(r-r_-)$, their Eqs.~(17) and (18) respectively. They also find
that the mass-density in between $r_-$ and $r_+$ is given in their
Eq.~(28), namely $\rho_{1 \rm o}=\frac{3}{8\pi} \left[
\frac{r_+-16\pi^2\sigma_{2 \rm e}^2r_-^3}
{r_+\left(r_+^2+r_+r_-+r_-^2\right)}\right]$, and the mass density
between $0$ and $r_-$ is de Sitter type. There is an interesting
special solution implicit in this solution. To find it
one abolishes the matter existent in between $r_-$ and $r_+$. Impose
$\rho_{1 \rm e}=0$ and $\rho_{1 \rm o}=0$. From $\rho_{1 \rm e}=0$
nothing special comes about, but from $\rho_{1 \rm o}=0$, one finds
that $\sigma_{2 \rm e}=\frac{1}{16\pi^2}\frac{r_+}{r_-^3}$.  Then this
solution is precisely our regular nonextremal black hole with a lightlike
matter boundary at the inner horizon $r_-$, as one can check (see our
part {\it (a)} of Section \ref{sec-pots}).

\vskip 0.1cm

\noindent(ii) The paper by Barrab\`es and Israel of 1991 \cite{barisrl91}
also finds the regular nonextremal black hole with a lightlike matter
boundary at the inner horizon $r_-$.

The direct matching conditions between an interior de Sitter region
and an exterior Reissner-Nordstr\"om region having a lightlike surface
as soldering surface was indeed considered by Barrab\`es and Israel
\cite{barisrl91} (see also \cite{barrabes_hogan}).  The authors have
noticed that different matching conditions at a horizon are possible,
and have given two special conditions, the static soldering and the
affine soldering. In general, the soldering can be performed in anyone
of the two horizons, but the surface pressure and energy density are
nonzero. However, the static soldering offers
a special case.
When the de Sitter radius $R$ is related to the charge of the
Reissner-Nordstr\"om solution $q$ by $R^2 = q^2/3$, then the surface
pressure and density are both zero and the surface gravities of the de
Sitter horizon and of the Cauchy Reissner-Nordstr\"om horizon are
equal.  Here, of course, the matching happens at the inner horizon.
This result agrees with what we have found above, but in our case, the
matching can be done at any surface $r=r_0$, timelike if $r_0 < r_-$,
or lightlike if $r_0=r_-$.  The solution is the same as the hidden
solution in Shen and Zhu \cite{sz51,sz52}, and so the same as our
particular solution.

\vskip 0.3cm
\centerline{\it (b) The solution is in the Weyl-Guilfoyle
class of solutions}

\noindent
Charged star-type solutions found by Guilfoyle \cite{guilfoyle}
have a plethora of parameters from which one can choose values. These
solutions contain quasiblack hole solutions as found
by Lemos and Zanchin \cite{lemoszanchin2010}
and it seems they also contain many different regular black holes.
Here, we indicate that in a particular limit of those charged star-type
solutions \cite{guilfoyle} one obtains the regular nonextremal black
holes with a lightlike matter boundary at the inner horizon mentioned in
Sec.~\ref{sec-pots}{\it (a)}.

Indeed, taking the parameter $A\to 0$ in Eq.~ (25) of \cite{guilfoyle},
or the limit $a\to\infty$ in \cite{lemoszanchin2010} as we do here, we
find
$
\displaystyle{\lim_{a\to\infty} \dfrac{q^2}{m^2}= \frac{1}{2}\left(2
+\left|\frac{r_0}{m} -2\right| - \frac{r_0}{m} \right)\frac{r_0}{m}}
$.
Due to the absolute value $\left|\frac{r_0}{m} -2\right|$ one concludes,
remarkably, there are two branches, $\dfrac{r_0}{m}>2$ and
$\dfrac{r_0}{m} < 2$. For  ${r_0}/{m}>2$ one finds
$
{\lim_{a\to\infty} \dfrac{q^2}{m^2}= 0}
$,
which gives the uncharged Schwarzschild interior solution,
and it is not of our concern here. For  ${r_0}/{m}<2$ one finds
$
{\lim_{a\to\infty} \dfrac{q^2}{m^2}=
\left(2 -\frac{r_0}{m}\right) \frac{r_0}{m}}
$,
which is equivalent to
$
1 -\frac{2m}{r_0}+\frac{q^2}{r_0^{2}}=0
$. 
This branch is the one that interests us here,
and we take the opportunity to make an analysis initiated in our
previous paper \cite{lemoszanchin2010} in relation to this branch. The
comments here take over the comments there.  Thus $r_0$ is at a
horizon, and consequently either at $r_+$ or $r_-$.  Under closer
scrutiny one finds that there is a special regular black hole solution
among a maze of regular black hole solutions, in which $r_0=r_-$,
$m=2r_0$, and $q=\sqrt3 r_0$.  This solution thus belongs to
Guilfoyle's class of solutions, i.e., it is a solution of Weyl-Guilfoyle
type.  Thus, this particular solution belongs to both Guilfoyle's
class of solutions and to a particular set of solutions we have been
studying here. The other regular black holes and stars we have found
here are not of Weyl-Guilfoyle type.

\section{Conclusions}
\label{sec-conclusion}

Charged regular black holes and overcharged stars as solutions to
certain distributions of
spherically symmetric charged matter have been displayed and studied.
The interior distribution of matter is constituted by a de
Sitter perfect fluid with pressure $p$ equal to the negative of the
energy density $\rho_{\rm m}$. Thus, the interior can be
interpreted as a false vacuum state. The interior metric is matched
into an exterior Reissner-Nordstr\"om electrovacuum
region. The Einstein-Maxwell equations together with the equation of
state $p=-\rho_{\rm m}$, imply that the isotropic pressure and the
energy density are constant throughout the spacetime, and that the
electric charge must be located at the boundary of the fluid
distribution.  Pressure and energy density both go to zero at the
boundary, and the resulting solutions represent regular charged black
holes.  The matter boundary is timelike, and in a limiting case is
lightlike. The boundary surface is always
at a radius smaller or, at most (in the lightlike case), equal to the
inner Reissner-Nordstr\"om horizon. For overcharged matter
a star pops up, and the horizons evanesce.

If the charge $q$ is the elementary charge $e$ then the mass of the
solution is of the order of the Planck mass and the radius is of the
order of the Planck radius. The solutions could then provide a model
for a charged elementary Planckian particle.

Due to the fact that the matter of the regular black hole is in the
inside or at the Cauchy horizon, these solutions may suffer the mass
inflation instability.

In some models, such as in Dymnikova's, the experimental investigation
of physical processes occurring near the external event horizon could
give some information about processes occurring deeply inside the
black hole.  Unfortunately, in our solution near the external event
horizon there is no way to probe the inside of the black hole.

\section*{Acknowledgments}
We thank conversations with Kirill Bronnikov and Oleg Zaslavskii.
This work was partially funded by Funda\c c\~ao para a Ci\^encia e
Tecnologia (FCT) - Portugal, through projects Nos.~CERN/FP/109276/2009
and PTDC/FIS/098962/2008.  JPSL thanks the FCT
grant SFRH/BSAB/987/2010.  VTZ thanks Funda\c c\~ao de Amparo \`a
Pesquisa do Estado de S\~ao Paulo (FAPESP) and Conselho Nacional de
Desenvolvimento Cient\'\i fico e Tecnol\'ogico of Brazil (CNPq) for
financial help. We thank Observat\'orio Nacional - Rio de Janeiro
for hospitality.

\appendix

\section{The matching conditions on the boundary}
\label{appA}

We will follow  \cite{israel} to show that the boundary at $r_{0}$ is
really
a boundary surface. We need to show that the metric, or first fundamental
form, $g_{\mu\nu}$ is continuous at the surface,
$g_{\mu\nu}^-=g_{\mu\nu}^+$, or what is the same thing,
$ds^2_-=ds^2_+$. It is also necessary that the
extrinsic
curvature, or second fundamental  form, $K_{\mu\nu}$ is continuous,
$K_{\mu\nu}^-=K_{\mu\nu}^+$.

For the surface $\Sigma$, defined by  $r=r_{0}$, we adopt the metric
\begin{equation}
ds_\Sigma^2=-d\tau^2+
r_0^2(d\theta^2+\sin^2 d\phi^2)
\,,
\label{metricoutersurface}
\end{equation}
with the intrinsic coordinates of $\Sigma$ being
$\xi^a=(\tau,\theta,\phi)$, while the inner and the outer metrics are 
given
receptively by (see Eqs.~(\ref{metricsphfinal})-(\ref{BAsol1-2})),
\begin{eqnarray}
ds_-^2& =& -\left(1-\frac{r^2}{R^2}\right)\,dt^2
+\left(1-\frac{r^2}{R^2}\right)^{-1}  dr^2\nonumber\\
& &  +\,r^2(d\theta^2+\sin^2
d\phi^2) \,,  \label{metric-}
\end{eqnarray}
and
\begin{eqnarray}
  ds_+^2&= &-\left(1-\frac{2m}{r}+
\frac{q^2}{r^2}\right)\,dt^2+ \left({1-\frac{2m}{r}+
\frac{q^2}{r}}\right)^{-1}{dr^2} \nonumber\\
& & + \,r^2(d\theta^2+\sin^2 d\phi^2)
\, , \label{metric+}
\end{eqnarray}
where we have identified the coordinates $(t,\,r,\,\theta,\,\varphi)$ in
both regions of the spacetime.

Let us assume that the boundary surface $\Sigma$ is timelike, which
means $1-r_0^2/R^2 > 0$ and also $1-2m/r_0 +q^2/r_0^2 >0 $.  Hence we
note first that, since the surface $\Sigma$ is spherical and
non-lightlike, the radial coordinate $r$ can be used as the matching
parameter along the generators on $\Sigma$, and so the normal $n_\mu$
to the surface has only the radial component
$n_r=\sqrt{g_{rr}}$. Therefore, in the present case the extrinsic
curvature has the form
\begin{equation}
K_{ab}^{\pm}=-n^{\pm}_{r} \Gamma ^{r (\pm)}_{\;\;\mu
\nu}\;\frac{\partial x^{\mu}}{\partial \xi ^{a}} \,
\frac{\partial x^{\nu}}{\partial \xi ^{b}}
\,,
\label{extrinsiccurvaturegeneral}
\end{equation}
where $\xi^a$ are the intrinsic coordinates of the surface.
Now we analyze the junction at the outer and inner surfaces.  The
continuity of the first fundamental form at the boundary implies that
$g_{tt}^+ = g_{tt}^- $ and $g_{rr}^+ = g_{rr}^- $. These two
conditions are satisfied by the continuity of the
line elements given in Eqs.~\eqref{metric-}-\eqref{metric+}
at $r=r_0$, namely,
\begin{equation}
1-r_0^2/R^2 = 1- 2m/r_0 + q^2/r_0^2 \,,
\label{matchg1}
\end{equation}
which is Eq.~\eqref{Rm2} and defines $R^{-2}$ in terms
of the parameters $r_0$, $m$, and $q$. One then
sees that the $g_{tt}^+$ and $g_{tt}^-$ match ant $r=r_0$, and
$g_{rr}^+$ and $g_{rr}^-$ also match, as well as the terms for the
angular part of the metric.  Now, at $r_{0}$ coming from the interior
one finds $K_{\tau\tau}^-=-n^{-}_r\, \Gamma ^{(-)r}_{\;\;tt}\,
\frac{dt}{d\tau}\frac{dt}{d\tau}$. By construction one has
$g^{\pm}_{tt}\,\dfrac{dt^2}{d\tau^2}=-1$. One also has
$n_r^-={1}/\sqrt{1-\frac{r_0^2}{R^2}}$. Noting that
$\dfrac{dg_{tt}^-}{dr}=- \dfrac{2r_0}{R^2}$ one finds $\Gamma
^{r(-)}_{\;\;tt}=\left(1-\dfrac{r_0^2}{R^2}\right)\dfrac{r_0}{R^2}$.
Thus,
$K_{\tau\tau}^-=
-\sqrt{1-\dfrac{r_0^2}{R^2}\,}\dfrac{r_0}{R^2}$. Similarly, the
extrinsic curvature coming from the exterior Reissner-Nordstr\"om
region can also be calculated. One finds
$K_{\tau\tau}^+=\sqrt{1-\dfrac{2m}{r_0}
+\dfrac{q^2}{r_0^2}}\left(\dfrac{m}{r_0^2} -
\dfrac{q^2}{r_0^3}\right)$. So, in order to match $K_{\tau\tau}^+$ and
$K_{\tau\tau}^-$ the parameters $R$, $r_0$, $m$ and $q$ must satisfy
the relation
\begin{equation}
\frac{r_0}{R^2}= -\frac{m}{r_0^2}+ \frac{q^2}{r_0^3} \,. \label{matchK}
\end{equation}
This result, together with Eq.~\eqref{matchg1}, gives
\begin{equation}
\frac{1}{R^2} = \frac{1}{3}\frac{q^2}{r_0^4} \,, \label{matchR}
\end{equation}
exactly the result found above (cf.~Eq.~\eqref{qtor0ratio}).
This also means
that the surface pressure vanishes.  One can easily check
that $K_{\theta\theta}$ and $K_{\phi\phi}$ also match.  So the metric
and the extrinsic curvature are continuous as is required. The
electric potential $\phi$ is also
continuous at the boundary as it is also required.

Let us assume that the boundary surface $\Sigma$ is lightlike now,
which means $1-r_0^2/R^2 > 0$ and also $1-2m/r_0 +q^2/r_0^2 >0 $.
Then, the normal vector $n^\mu$ is lightlike. The radial coordinate
$r$ cannot be used for the matching in this case. As it is proposed in
\cite{barrabes_hogan} (see also \cite{barisrl91} for the original
work), an alternative is using the advanced (retarded) time $u$ as the
soldering parameter, so that we can perform what Barrab\`es and Israel
\cite{barisrl91} call a static soldering. The new coordinate $u$ is
defined by
\begin{eqnarray}
du &=& dt +\epsilon \frac{dr}{\left({1-\dfrac{r^2}{R^2}}\right)}\, ,\quad
\mbox{for } r< r_0, \\
du &=& dt +\epsilon \frac{dr}{\left({1-\dfrac{2m}{r}+\dfrac{q^2}{r^2}}
\right)}\,,\quad, \mbox{for}\; r> r_0,
\end{eqnarray}
where $\epsilon = \pm 1$, and the plus (minus) sign is associated to the
outgoing (ingoing) radial light rays on the cone $u=$ constant.
The parameters on $\Sigma$ may be chosen as
$\xi^a=(\tau=u,\,\theta,\,\varphi)$ \cite{barrabes_hogan,barisrl91}.
The normal to $\Sigma$ is therefore $n^\mu = \dfrac{\partial 
x^\mu}{\partial
u}$, and, moreover, since $n^\mu$ is lightlike, an additional
transverse vector $N^\mu$ is needed. This can be chosen as $N^\mu
=\epsilon\dfrac{\partial x^\mu}{\partial r} $. Instead of the extrinsic
curvature $K_{ab}$, the relevant quantity for a lightlike soldering
is the transverse curvature ${\cal K}_{ab}$ defined by
\beq
{\cal K}_{ab}= - N_\mu \Gamma^\mu_{\nu\rho}\frac{\partial
x^{\nu}}{\partial \xi ^{a}} \,\,
\frac{\partial x^{\rho}}{\partial \xi ^{b}}.
\eeq
Then, the matching of the first fundamental form at $r=r_0$ gives 
$g_{uu}^+
= g_{uu}^-$, which results in Eq.~\eqref{matchg1}, is accomplished
as far as that equality holds even in the limit of $r_0$ being a
solution 
of
the equations $1-r_0^2/R^2 = 0$ and  $1- 2m/r_0 + q^2/r_0^2 =0$, which is
true in the particular case we are considering here.
Moreover, calculating the nonzero components
of ${\cal K}^{\pm}_{ab}$ we get
\beqa
{\cal K}^{+}_{uu} = \epsilon\left(\frac{m}{r_0^2}
-\frac{q^2}{r_0^3}\right)\,,\label{Kuu_ext}\\
{\cal K}^{+}_{\theta\theta}=  {\cal
K}^{+}_{\varphi\varphi}\sin^{-2}\theta
=
{\epsilon}{r_0},
\eeqa
and
\beqa
{\cal K}^{-}_{uu} =  -\epsilon\frac{r_0}{R^2}\,,\label{Kuu_int}\\
{\cal K}^{-}_{\theta\theta}=  {\cal
K}^{-}_{\varphi\varphi}\sin^{-2}\theta
=
{\epsilon}{r_0},
\eeqa
where $r_0$ coincides with one of the horizons of the
Reissner-Nordstr\"om
spacetime. The matching between the metric coefficients on the horizon
$r_0$
is clear. Again, imposing the continuity of the $uu$ component of
transverse
curvature, i.e., equating the relations~(\ref{Kuu_ext}) and 
(\ref{Kuu_int})
we find that they match only in the case $R=r_0=r_-$, $r_-$ being the 
inner
Reissner-Nordstr\"om horizon. As a consequence of this matching, the 
pressure
vanishes at the boundary surface, $p(r_0)=0$.
Note that here the lightlike case can be also analyzed as a
limit of the configurations with a timelike boundary.

\section{How to obtain $p^\prime$}
\label{appB}

Note that the equation
$\theta^2(r-r_0)=\theta(r-r_0)$ can only be used after
taking derivatives. Also the Dirac delta function is
given by $\delta(r-r_0)=[\theta(r-r_0)]^{\,\prime}$.
Thus in Eq.~(\ref{pressure02}) the term in front
of $\dfrac{q^2}{r_0^4}$ differentiates to
$\left[\theta^2(r-r_0)\theta(r_0-r)\right]^{\,\prime}=
2\theta(r-r_0)\delta(r-r_0)\theta(r_0-r)
-\theta^2(r-r_0)\delta(r-r_0)
=2\delta(r-r_0)-\theta(r-r_0)\delta(r-r_0)=\delta(r-r_0)
$, where we have used $\theta(r-r_0)\theta(r_0-r)=1$ in $r=r_0$
and zero otherwise, and
$\theta(r-r_0)\delta(r-r_0)=\delta(r-r_0)$. Now, the term in front of
$-\dfrac{3}{R^2}$ has derivative
$[\theta(r_0-r)]^{\,\prime}=-\delta(r-r_0)$. Thus
the whole term yields  $+\dfrac{3}{R^2}\delta(r-r_0)$.
Since $\dfrac{3}{R^2}=\dfrac{q^2}{r_0^4}$ we have to sum
both terms to give Eq.~(\ref{pressureprime}),
i.e., $p^{\,\prime}(r)=\frac{q^2}{4\,\pi\,r_0^4}\,\delta(r-r_0)$.

\end{document}